\documentclass[journal]{IEEEtran}
\ifCLASSINFOpdf
\else
\fi
\usepackage{mathtools}
\usepackage{color}
\usepackage{algorithm}
\usepackage[noend]{algpseudocode}

\usepackage{times}
\usepackage{epsfig}
\usepackage{graphicx}
\usepackage{amssymb}
\usepackage{romannum}
\usepackage{color}

\usepackage{amsmath}
\usepackage{breqn}
\usepackage{multirow}
\usepackage{subfigure}
\usepackage{xurl}
\usepackage{caption}
\usepackage{changepage}
\usepackage{textcomp,marvosym}

\begin{document}
%
% paper title
% Titles are generally capitalized except for words such as a, an, and, as,
% at, but, by, for, in, nor, of, on, or, the, to and up, which are usually
% not capitalized unless they are the first or last word of the title.
% Linebreaks \\ can be used within to get better formatting as desired.
% Do not put math or special symbols in the title.
\title{Enhancing Deep Knowledge Tracing via Diffusion Models for Personalized Adaptive Learning}
%
%
% author names and IEEE memberships
% note positions of commas and nonbreaking spaces ( ~ ) LaTeX will not break
% a structure at a ~ so this keeps an author's name from being broken across
% two lines.
% use \thanks{} to gain access to the first footnote area
% a separate \thanks must be used for each paragraph as LaTeX2e's \thanks
% was not built to handle multiple paragraphs
%

 \author{Ming Kuo,~Shouvon Sarker,~Lijun Qian,~Yujian Fu, ~Xiangfang Li,~Xishuang Dong
        % <-this % stops a space
\thanks{M. Kuo, S. Sarker, L. Qian, X. Li and X. Dong are with the Department of Electrical and Computer Engineering, Prairie View A\&M University, Texas A\&M University System, Prairie View, TX 77446, USA. Y. Fu is with Department of Electrical Engineering and Computer Science,
School of Engineering, Technology \& Physical Sciences, Alabama A\&M University, Madison, Alabama, USA. Email: mmkuo@pvamu.edu, ssarker3@pvamu.edu, liqian@pvamu.edu, yujian.fu@aamu.edu, xili@pvamu.edu, xidong@pvamu.edu}% <-this % stops a space
%\thanks{J. Doe and J. Doe are with Anonymous University.}% <-this % stops a space
%\thanks{Manuscript received April 19, 2005; revised August 26, 2015.}
}

\maketitle

% As a general rule, do not put math, special symbols or citations
% in the abstract or keywords.
\begin{abstract}

In contrast to pedagogies like evidence-based teaching, personalized adaptive learning (PAL) distinguishes itself by closely monitoring the progress of individual students and tailoring the learning path to their unique knowledge and requirements.  A crucial technique for effective PAL implementation is knowledge tracing, which models students' evolving knowledge to predict their future performance. Based on these predictions, personalized recommendations for resources and learning paths can be made to meet individual needs. Recent advancements in deep learning have successfully enhanced knowledge tracking through Deep Knowledge Tracing (DKT). This paper introduces generative AI models to further enhance DKT. Generative AI models, rooted in deep learning, are trained to generate synthetic data, addressing data scarcity challenges in various applications across fields such as natural language processing (NLP) and computer vision (CV). This study aims to tackle data shortage issues in student learning records to enhance DKT performance for PAL. Specifically, it employs TabDDPM, a diffusion model, to generate synthetic educational records to augment training data for enhancing DKT. The proposed method's effectiveness is validated through extensive experiments on ASSISTments datasets. The experimental results demonstrate that the AI-generated data by TabDDPM significantly improves DKT performance, particularly in scenarios with small data for training and large data for testing. 

\end{abstract}

% Note that keywords are not normally used for peerreview papers.
\begin{IEEEkeywords}
Personalized Adaptive Learning, Deep Knowledge Tracing, Generative AI Models, Diffusion Models
\end{IEEEkeywords}

% For peer review papers, you can put extra information on the cover
% page as needed:
% \ifCLASSOPTIONpeerreview
% \begin{center} \bfseries EDICS Category: 3-BBND \end{center}
% \fi
%
% For peerreview papers, this IEEEtran command inserts a page break and
% creates the second title. It will be ignored for other modes.
\IEEEpeerreviewmaketitle

\section{Introduction }
\label{sec1}

%\section{Introduction}

Unlike various teaching methods such as evidence-based teaching, personalized adaptive learning (PAL)~\cite{peng2019personalized} stands out by continuously monitoring each student's progress and adjusting the educational journey according to their distinct knowledge and needs. Instead of a one-size-fits-all approach, PAL tailors the learning journey for every learner. A pivotal tool in making PAL effective is knowledge tracing (KT)~\cite{abdelrahman2023knowledge}, which tracks the progression of students' understanding to anticipate their upcoming performance. Using these insights, tailored suggestions for study materials and pathways can be provided to cater to individual requirements. Additionally, content that might be deemed too elementary or overly advanced can be bypassed or postponed.

Recent progress in deep learning has notably improved KT with the introduction of Deep Knowledge Tracing (DKT)~\cite{piech2015deep}. DKT is designed to leverage advanced deep learning methods to understand the temporal patterns in sequences of student interactions, predicting a student's response to a new question. Previous work has demonstrated that DKT surpasses conventional KT models across various benchmark datasets. The evolution of DKT has been influenced by diverse approaches, including memory architectures~\cite{abdelrahman2019knowledge}, attention mechanisms~\cite{ghosh2020context}, graph-based learning~\cite{nakagawa2019graph}, textual features~\cite{liu2019ekt}, and forgetting  forgetting~\cite{abdelrahman2022deep}. Furthermore, some research work~\cite{lee2022contrastive, song2022bi, lee2021consistency} have delved into data augmentation strategies to boost DKT's efficiency. Nevertheless, there's limited exploration into harnessing Generative AI methods~\cite{cao2023comprehensive} to bolster DKT by generating synthetic data.

Generative AI (GAI) models, as explored by Cao et al.\cite{cao2023comprehensive}, leverage deep learning techniques to create synthetic data, effectively addressing data scarcity challenges in diverse applications like natural language processing (NLP) and computer vision (CV). Key players in the realm of GAI models include ChatGPT and DALL-E-2\cite{ramesh2021zero}, both rooted in transformer architecture, featuring both an encoder and a decoder~\cite{qiu2020pre}. These pretrained GAI models extend their capabilities beyond initial training objectives by excelling in downstream tasks. A notable advancement in the field is the emergence of diffusion models~\cite{song2019generative, dhariwal2021diffusion} as the latest state-of-the-art GAI models. This family of models has disrupted the longstanding dominance of generative adversarial networks (GANs)~\cite{saxena2021generative} in the intricate task of image synthesis~\cite{song2020score}. Moreover, diffusion models exhibit promise across various domains, such as computer vision~\cite{yang2022lossy, yang2023diffusion} and natural language processing~\cite{li2022diffusion}.

This paper introduces diffusion models as a solution to address data scarcity challenges in student learning records, aiming to enhance the performance of DKT for PAL.  In particular, the task involves small data for training and large data for testing. In detail, the proposed approach utilizes TabDDPM~\cite{kotelnikov2023tabddpm}, a diffusion model capable of generating synthetic educational records. TabDDPM exhibits universality, applicable to any tabular dataset and capable of handling various feature types. Notably, it outperforms alternative methods designed for tabular data, including GAN-based and VAE-based approaches. The proposed method’s effectiveness is validated through extensive experiments on ASSISTments datasets. The experimental results demonstrate that the AI-generated data by TabDDPM significantly improves DKT performance, particularly in scenarios with small data for training and large data for testing.

The contributions in this paper can be summaries as:

\begin{enumerate}
\item 
To the best of our knowledge, this study represents the pioneering exploration of resolving data shortage issues on DKT through GAI models. The approach leverages diffusion models to generate educational records, effectively augmenting the original training datasets used to train DKT models. 

\item The experimental results, evaluated using various metrics, unequivocally show a significant improvement in DKT performance through the proposed method. Additionally, there is a consistent increase in DKT performance as the amount of AI-generated training data increases. This finding aligns with previous research on data augmentation through Generative AI (GAI) techniques, further validating the effectiveness of leveraging AI for enhancing the training datasets in educational contexts.

\end{enumerate}

\section{Methodology}
\label{sec2}
% Methodology

This paper aims  to address the data scarcity issue in Dynamic Knowledge Tracing (DKT) by generating synthetic data using diffusion models. The approach encompasses the implementation of DKT models and the utilization of specific diffusion models for synthetic data generation.

\subsection{Deep Knowledge Tracing (DKT)}
Inspired by the groundbreaking advancements in deep learning~\cite{goodfellow2016deep}, there has been swift progress in the development of knowledge tracing models. A notable contribution in this domain is Deep Knowledge Tracing (DKT)~\cite{piech2015deep}, which leverages Recurrent Neural Networks (RNNs)~\cite{lipton2015critical} to capture temporal dynamics within a sequence of interactions between a student's questions and answers. This enables the model to predict a student's response to a new question based on their historical interactions. A fundamental representation of a simple RNN network for DKT is defined as follows:

\begin{equation}
h_{t} = tanh(W_{hx}x_{t} + W_{hh}h_{t-1} + b_{h})
\end{equation}

\begin{equation}
y_{t} = \sigma(W_{yh}h_{t} +  b_{y})
\end{equation}

where both $tanh(\cdot)$ and $\sigma(\cdot)$ are applied element-wise. The model is parameterized by an input weight matrix $W_{hx}$, a recurrent weight matrix $W_{hh}$, an initial state $h_{0}$, and a readout weight matrix $W_{yh}$. Biases for the latent and readout units are denoted by $b_{h}$ and $b_{h}$. The inputs ($x_{t}$) to the dynamic network can be either one-hot encodings or compressed representations of a student's action, while the prediction ($y_{t}$) is a vector representing the probability of correctly answering each exercise in the dataset. Empirical results have demonstrated that DKT surpasses traditional KT models on various benchmark datasets. This endeavor underscores the potential of employing deep learning models to effectively address the KT problem~\cite{abdelrahman2023knowledge}.

\subsection{TabDDPM}

Denoising diffusion probabilistic models (DDPM)~\cite{ho2020denoising} have garnered significant attention in the generative modeling community, often surpassing alternative methods in producing both realistic and diverse samples~\cite{dhariwal2021diffusion}. Kotelnikov et al. introduced TabDDPM~\cite{kotelnikov2023tabddpm} to explore the applicability of DDPM's versatility to general tabular challenges. Such challenges are prevalent in numerous industrial applications characterized by datasets comprising diverse and heterogeneous features. TabDDPM offers a streamlined DDPM variant tailored for tabular data, accommodating mixed data types, including numerical and categorical features. For instance, a tabular data sample is structured as $x=[x_{num}, x_{cat_{1}}, ..., x_{cat_{C}}]$, encompassing $x_{num}$ numerical features and $C$ categorical features denoted by $x_{cat_{i}}$. Notably, TabDDPM employs multinomial diffusion to represent categorical and binary features, while Gaussian diffusion is employed for numerical features. The reverse diffusion process within TabDDPM is characterized by a multi-layer neural network, producing outputs equivalent in dimensionality to $x_{0}$. During the network's training, the objective encompasses minimizing a combined loss function, comprising the mean-squared error $L_{t}^{simple}$ for the Gaussian diffusion  to process numerical attributes of tabular data and individual KL divergences $L_{t}^{i}$ for each multinomial diffusion to process categorial attributes of tabular data. Furthermore, the cumulative loss for multinomial diffusions is adjusted by dividing it by the count of categorical features, ensuring comprehensive training dynamics.

\begin{equation}
L_{t}^{TabDDPM} = L_{t}^{simple} + \frac{\sum_{i \leq C}L_{t}^{i} }{C}
\end{equation}

\subsection{Proposed Method}

\begin{figure} [ht]
 	\centering
	\includegraphics[width=1.\linewidth]{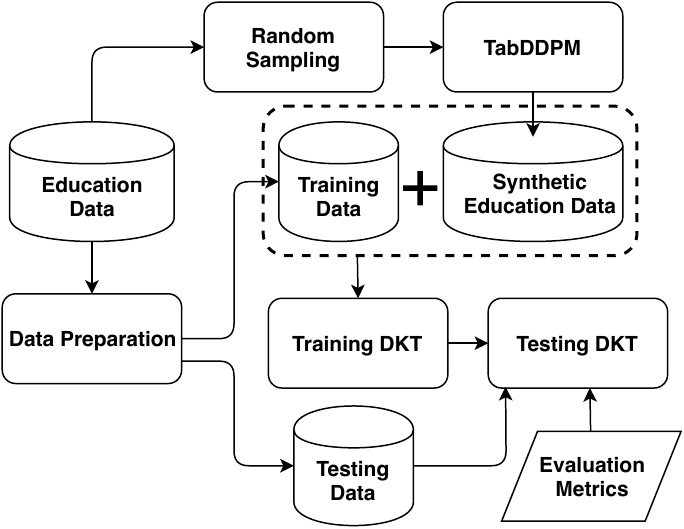}
	\caption{Flow of the proposed method.}
	\label{Fig_model}
\end{figure}

The flow of the proposed method is depicted in Figure~\ref{Fig_model}. The process begins by randomly sampling data, serving as inputs to TabDDPM, which generates synthetic education data. Subsequently, this synthetic education data is integrated with the original training education data, forming a combined dataset used to train the DKT model for knowledge tracing. Finally, the trained DKT model is evaluated using testing education data, employing multiple evaluation metrics to assess its performance.

\section{Experiment }
\label{sec3}

\subsection{Dataset}

\begin{table*}[!ht]
\begin{adjustwidth}{0in}{0in}
	\caption{ Comparing performance between DKT and the proposed method (DKT + TabDDPM). }
       
        \begin{center}
                \begin{tabular}{|l|cccc|}
                \hline \textbf{Model} 				& \textbf{Accuracy (\%)} & \textbf{AUC (\%)} & \textbf{Precision  (\%)} & \textbf{Recall  (\%)}  \\ \hline 
                		  	DKT					&  58.38\textpm0.98				& 54.59\textpm1.27				& 60.45\textpm0.56 				& 86.04\textpm3.70			\\ 
                    	        DKT + TabDDPM		&  \textbf{63.46\textpm0.48}				& \textbf{66.00\textpm0.85}		& \textbf{63.08\textpm0.44} 				& \textbf{92.40\textpm1.06}		 	\\
								 \hline
                \end{tabular}
       \end{center}
         \label{tab_cp}
         \end{adjustwidth}
\end{table*}

\begin{table*}[!ht]
\begin{adjustwidth}{0in}{0in}
	\caption{ Comparing performance between various proposed method. }
       
        \begin{center}
                \begin{tabular}{|l|cccc|}
                \hline \textbf{Model} 				& \textbf{Accuracy (\%)} & \textbf{AUC (\%)} & \textbf{Precision  (\%)} & \textbf{Recall  (\%)}  \\ \hline 
                	
                    	        DKT + TabDDPM ($1,000$)		&  59.30\textpm0.43				&  57.29\textpm1.14				&   60.50\textpm0.16				&   90.10\textpm2.50		 	\\
	         		DKT + TabDDPM ($3,000$)		&  60.72\textpm0.95				& 62.10\textpm0.52				& 61.14\textpm0.52 				& 90.60\textpm1.15 		 	\\
			 	DKT + TabDDPM ($5,000$)		&  60.22\textpm2.22				& 60.02\textpm4.48				& 61.51\textpm0.78 				& 87.55\textpm6.14		 	\\
			  	DKT + TabDDPM ($7,000$)		&  62.23\textpm0.43				& 63.78\textpm0.77				& 62.42\textpm0.26				& 91.01\textpm1.60		 	\\
			   	DKT + TabDDPM ($10,000$)		&  62.23\textpm0.43				& 63.78\textpm0.77				& 62.42\textpm0.26 				& 91.01\textpm1.60	 	\\
			    	DKT + TabDDPM ($15,000$)		&  \textbf{63.62\textpm0.32}		& 65.97\textpm0.43				& \textbf{63.45\textpm0.33} 				& 90.97\textpm0.95	 	\\
				DKT + TabDDPM ($30,000$)		&  63.46\textpm0.48				& \textbf{66.00\textpm0.85}		& 63.08\textpm0.44 				& 92.40\textpm1.06 		 	\\
				DKT + TabDDPM ($50,000$)		&  63.46\textpm0.44				& 65.63\textpm0.77				& 63.17\textpm0.36				& 91.85\textpm0.66 		 	\\
				DKT + TabDDPM ($100,000$)		&  63.44\textpm0.39				&  65.63\textpm0.71 				&  62.95\textpm0.25		&  \textbf{92.97\textpm0.62} 		 	\\
	          
								 \hline
                \end{tabular}
       \end{center}
         \label{tab_cp_2}
         \end{adjustwidth}
\end{table*}

The ASSISTments datasets\footnote{https://sites.google.com/site/assistmentsdata/} consist of longitudinal data gathered from the ASSISTment platform\footnote{https://new.assistments.org/individual-resource/skillbuilders}, an online tutoring service. Widely recognized as the predominant datasets for evaluating Knowledge Tracing (KT) models, they feature the highest number of questions overall. These datasets comprise math exercises from grade school, sourced from the Massachusetts Comprehensive Assessment System (MCAS), encompassing various question types like multiple choice, text, and open-ended questions. Multiple versions of the ASSISTments datasets exist, each collected during distinct time periods. For this study, ASSISTments2009 was utilized, collected during the 2009-2010 school year, initially containing $525,535$ interactions (i.e., student responses to questions in the dataset), including duplicates. 

In real-world applications, the amount of training data often falls short compared to the testing data used by users to apply AI models for specific tasks. To address this need, this study splits the ASSISTments2009 dataset into training data (5\%), validation data (20\%), and testing data (75\%). To generate synthetic educational data, TabDDPM~\cite{kotelnikov2023tabddpm} is applied to the ASSISTments2009 data, resulting in multiple education datasets. Only two categorical attributes, namely \textit{skill id} and \textit{user id}, along with one continuous attribute \textit{overlap_time}, and the ground truth \textit{correct} are utilized in this process.

\subsection{Evaluation metrics}

Various evaluation metrics are employed to evaluate the performance of our proposed model, which includes Accuracy, precision, and recall. 

Accuracy is calculated by dividing the number of student answers identified correctly over the total number of student answers. 

\begin{equation}
	Accuracy = \frac{N_{correct}}{N_{total}}.
\end{equation}

$Precision$ defines the capability of a model to represent only correct student answers while $Recall$ computes the aptness to refer all corresponding correct student answers.

\begin{equation}
	Precision = \frac{TP}{TP+FP}.
\end{equation}

\begin{equation}
	Recall = \frac{TP}{TP+FN}.
\end{equation}
where ${TP}$ (True Positive) tallies the total number of student answers that match the ground truth, ${FP}$ (False Positive) quantifies the number of student answers that do not align with the ground truth but are incorrectly identified as correct. ${FN}$ (False Negative) enumerates the instances where correct student answers are incorrectly classified as incorrect. Additionally, the study utilizes the Area Under the Curve (AUC) as a metric, which gauges the binary classifier's efficacy in distinguishing between classes. A higher AUC indicates superior performance in discerning between positive and negative classes.

\subsection{Experimental results and discussion}

Table~\ref{tab_cp} illustrates a comparison of performance between the baseline model (DKT) and the proposed model. The proposed model surpasses DKT in terms of accuracy, AUC, precision, and recall. Notably, TabDDPM-generated samples contribute significantly to enhancing AUC, recall, and accuracy, with a more pronounced impact on reducing false negatives to increase the recall.

Additionally, Table~\ref{tab_cp_2} provides a detailed breakdown of performance based on varying numbers of generated samples. Notably, even a small number of samples, such as $1,000$ samples, contributes to improving knowledge tracing (KT) performance. As the number of generated samples increases, the proposed method exhibits increased stability, reflected in the standard deviation of accuracy, precision, and recall. This aligns with prior research suggesting that involving more data in the training process enhances model performance.

\begin{figure} [ht]
 	\centering
	\includegraphics[width=1.\linewidth]{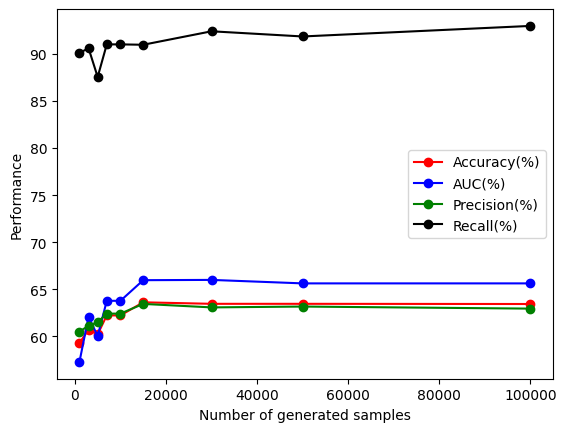}
	\caption{Performance comparison.}
	\label{Fig_result}
\end{figure}

Furthermore, Figure~\ref{Fig_result} visually depicts that an increased amount of generated data further enhances performance. This observation aligns with the findings presented in Table~\ref{tab_cp_2}, emphasizing the positive impact of augmented data on model performance.

\section{Related Work }
\label{sec4}

\subsubsection{Knowledge Tracing}

Knowledge Tracing (KT) is designed to effectively monitor a student's learning progress as they interact with teaching materials~\cite{abdelrahman2023knowledge}. In essence, KT aims to observe, represent, and quantify a student's knowledge state, including their mastery level of the skills underlying the teaching materials. It can be broadly categorized into traditional knowledge tracing and deep learning-based knowledge tracing. Traditionally, two prominent work for knowledge tracing are Bayesian Knowledge Tracing~\cite{cui2019analyzing} and Factor Analysis Models~\cite{yeung2019deep}. Conversely, recent advancements in deep learning have inspired researchers to apply deep learning techniques to knowledge tracing. In this context, the knowledge tracing task is commonly framed as a sequence prediction task with machine learning. Recurrent Neural Network (RNN) based techniques have significantly advanced the development of deep knowledge (DK) in the context of knowledge tracing~\cite{xiong2016going}. Despite these advancements, a persistent challenge in enhancing KT for real-world applications is the limited availability of education data. The scarcity of such data poses an ongoing obstacle to further improving the effectiveness of knowledge tracing methodologies in practical educational settings.

\subsubsection{Generative Models for Tabular Data Generation}

The exploration of generative models for tabular data generation is currently a vibrant area of research due to the increasing demand for high-quality synthetic data in various tabular tasks. This demand arises for several reasons. Firstly, tabular datasets, especially in the education domain, are often limited in size compared to datasets in vision or natural language processing (NLP) problems, which can leverage massive amounts of ``extra" data available from the Internet. Secondly, creating synthetic datasets that do not include actual user data allows for public sharing without compromising anonymity. In addition, unlike unstructured images or natural texts, tabular data is typically structured, and the interpretability of individual features raises questions about the necessity for complex ``deep" architectures. Therefore, simple interpolation techniques such as SMOTE~\cite{chawla2002smote} (originally designed for addressing class imbalance) can serve as effective solutions.

Recent research has introduced various models for tabular data generation, including tabular Variational Autoencoders (VAEs)~\cite{xu2019modeling}, GAN-based approaches~\cite{torfi2022differentially}, and diffusion model-based approaches~\cite{kotelnikov2023tabddpm}. These models contribute to addressing the challenges associated with limited tabular data size and the need for synthetic datasets in a privacy-preserving manner.

\section{Conclusion and Future Work}
\label{sec5}

Knowledge tracing stands as a crucial technique for implementing personalized adaptive learning. In this paper, diffusion models were utilized to improve deep knowledge tracing, with a specific focus on applying TabDDPM to generate synthetic education data. This augmentation aims to address the limitations posed by the scarcity of training data for real-world knowledge tracing applications.

In the future, we envisions exploring two promising directions for future research. Firstly, there is a plan to extend the generation of synthetic education data to encompass more aspects of knowledge tracing tasks. This expansion aims to enhance the diversity and richness of the generated data, contributing to a more comprehensive understanding of the learning process. Secondly, the proposed model will be validated in the context of personalized learning path recommendation to further advance personalized adaptive learning. This direction emphasizes tailoring educational experiences based on individual learning patterns, fostering a more effective and personalized approach to learning. The combination of these future directions is anticipated to contribute to the ongoing evolution of knowledge tracing and its applications in the personalized adaptive learning scenario. 

% use section* for acknowledgment
\section*{Acknowledgment}
\label{acknowledgement}
This research work is supported by NSF under award number 2235731. Any opinions, findings, and conclusions or recommendations expressed in this work are those of the author(s) and do not necessarily reflect the views of NFS.

% Can use something like this to put references on a page
% by themselves when using endfloat and the captionsoff option.
\ifCLASSOPTIONcaptionsoff
  \newpage
\fi

% trigger a \newpage just before the given reference
% number - used to balance the columns on the last page
% adjust value as needed - may need to be readjusted if
% the document is modified later
%\IEEEtriggeratref{8}
% The "triggered" command can be changed if desired:
%\IEEEtriggercmd{\enlargethispage{-5in}}

% references section
\bibliographystyle{IEEEtran}
\bibliography{Reference}

\end{document}